\documentstyle[12pt]{article}
\newcommand{\be}{\begin{equation}}
\newcommand{\ee}{\end{equation}}
\newcommand{\bd}{\begin{displaymath}}
\newcommand{\ed}{\end{displaymath}}
\newcommand{\baa}{\begin{array}{lll}}
\newcommand{\eaa}{\end{array}}
\newcommand{\ba}{\begin{eqnarray}}
\newcommand{\ea}{\end{eqnarray}}

\begin{document}
\begin{flushright}
CPT--99/P.3845.
\end{flushright}

\begin{center}
{\Large\bf Improved positivity bound for Deep Inelastic Scattering 
on transversely polarized nucleon} \\[10mm]

{\bf \large J. Soffer \footnote{E-mail: soffer@cpt.univ-mrs.fr
}}\\

{\it Centre de Physique Th\'eorique - CNRS - Luminy,\\
Case 907 F-13288 Marseille Cedex 9 - France} \\

{ \bf \large and \ \ O. V. Teryaev\footnote{
E-mail: teryaev@thsun1.jinr.ru}} \\

{\it Bogoliubov Laboratory of Theoretical Physics, \\
Joint Institute for Nuclear Research, Dubna, 141980, Russia}
\end{center}
\begin{abstract}
The positivity bound for the transverse asymmetry $A_2$ may be improved
by making use of the fact, that the state  
of a photon and a nucleon with total spin 3/2, does 
 not participate to the interference.
The bound is therefore useful in the case of a longitudinal asymmetry small (say, at low $x$)
or negative (like in the neuteron case). 

\end{abstract}


Positivity is playing a very important role in constraining various spin- \\
dependent observables, in particular by providing a bound for the 
transverse asymmetry in polarized Deep Inelastic Scattering (DIS). 
It is a well-known condition established long time ago and based on
an extensive study by
Doncel and de Rafael \cite{DDR}, written in the form
\begin{eqnarray}
\label{DDR}
|A_2| \leq \sqrt{R}~,
\end{eqnarray}
where  $A_2$ is the usual transverse asymmetry and
$R=\sigma_L/\sigma_T$ is the standard ratio in DIS of the cross section of 
longitudinally to transversely polarized off-shell photons. 
It reflects a non-trivial
positivity condition one has on the photon-nucleon helicity amplitudes. 
By substituting photons for gluons, we found earlier\cite{ST}, that 
the similar bound holds for the various matrix elements for 
longitudinal gluons in a nucleon \cite{GI} 
\begin{eqnarray}
\label{ineq}
|\Delta G_T(x)| \leq \sqrt{1/2G(x)G_L(x)}~.
\end{eqnarray}
However, this bound can be rederived in line with the positivity bound in the
quark case, known as Soffer inequality \cite{S}, 
 \begin{eqnarray}
|h_1(x)| \leq q_+(x) = {1 \over 2} [q(x) + \Delta q(x)]~,
\label{S}\end{eqnarray}
by making the substitution in Eq.(2), $G(x) \to G_+(x)
= {1 \over 2} [G(x) + \Delta G(x)]$, and providing a stronger 
restriction, especially when the gluon helicity distribution $\Delta G(x)$
is small or even negative.  
Coming back to the photon case, if $A_1$ denotes the asymmetry with longitudinally
polarized nucleon, we are led to 
\begin{eqnarray}
\label{DDRn}
|A_2| \leq \sqrt{R(1+A_1)/2}~,
\end{eqnarray}
a stronger bound than Eq.(1).
In the present paper we will show that this is really the case,
using a transparent physical approach, and we will comment on,
why, we think the weaker bound was used up to now. 

We start with the following expressions for 
the various photon-nucleon cross-sections in terms of the 
matrix elements describing the transition from the state $|H,h>$ of a
nucleon with helicity $h$ and a photon with helicity $H$, to the unobserved state 
 $|X>$
\begin{eqnarray}
\label{def}
\sigma^{\pm}_T=\sum_{X} |<+1/2, +1|X>|^2 \pm |<+1/2,
-1|X>|^2~,\nonumber \\
\sigma_L=\sum_{X} |<+1/2, 0|X>|^2=\sum_{X}
|<-1/2,0|X>|^2~, 
\nonumber \\
\sigma_{LT}=
2 Re \sum_{X} <+1/2,
+1|X> <-1/2, 0|X>~.
\end{eqnarray}
Note that while longitudinal and transverse cross-sections
are symmetric with 
respect to the reverse of the nucleon and photon helicities, this
is not the case for the interference term. The reason is very simple:
the opposite helicities of photon and nucleon correspond to their spins parallel
, so that the angular momentum of the state  $|X>$ has its maximum value 
$3/2$. The amplitude, which could  possibly interfere with it
to produce the transverse asymmetry, should have the same total
angular momentum of the state  $|X>$.
This is however impossible, as 
the flip of the one of the helicities
would require another one to exceed its maximal possible value,
in order to keep the angular momentum of $|X>$ the same. 
Therefore the interference, responsible for $A_2$, does not occur. 
This is quite a general reason,
for the occurence of the $+$ helicity configurations 
in all the cases considered above. 

We are now ready to write down the Cauchy-Schwarz inequality as
\begin{eqnarray}
\label{CS}
\sum_{X}|<+1/2,+1|X> \pm a<-1/2,0|X>|^2 \geq 0~,
\end{eqnarray}
where $a$ is a positive real number. 
By making use of the definitions (\ref{def}) and
after the standard minimization with respect to the choice of
$a$, one immediately arrives at
\begin{eqnarray}
\label{DDRs}
|\sigma_{LT}| \leq \sqrt{\sigma_L \sigma^+_T}~, 
\end{eqnarray}
leading directly to (\ref{DDRn}). 
The use of the new bound is resolving partially the puzzle, 
why the measured $A_2$ 
is such a small quantity. The fact, that the bound (1) is far
from being saturated is obvious at low $x$ in the proton
case, because, according to (4), it should be 
decreased by a factor $\sqrt 2$ due to the small longitudinal asymmetry.  
The bound under consideration is even more useful with a negative 
longitudinal asymmetry, like in the neutron case. 
Recall that for a pure 3/2 configuration we have $A_1=-1$ which implies 
$A_2=0$

One should note finally,
that this result is actually coming from the original papers
\cite{CL,DDR}, while it was somehow  weakened
and transformed to a more 
suitable form Eq.(1), because one was willing to exclude 
$A_1$ which was poorly known twenty years ago. To be convince of that,
 one should look at Eq.(2.40a) in
\cite{DDR}, which was, in fact, already contained in \cite{CL}. 

To conclude, we rederived a known, but so far forgotten stronger bound
for the transverse asymmetry in polarized DIS. 

We are indebted to E. de Rafael for useful dicsussions and to
Z-E. Meziani and R. Windmolders for interest in the work.


\begin{thebibliography}{99}


\bibitem{DDR} M.G. Doncel and E. de Rafael,
Nuovo Cimento {\bf 4A}, 363 (1971)


\bibitem{ST} J.Soffer and O.V.Teryaev, Phys.Lett. {\bf B419}, 400 (1998)


\bibitem{GI} A.S.Gorsky and B.L.Ioffe, Particle World vol.1,114 (1990).


\bibitem{S}
        J. Soffer, Phys. Rev. Lett. {\bf 74} (1995) 1292.

\bibitem{CL} N. Christ and T.D. Lee,  Phys.Rev. {\bf 143}, 1310 (1966).



\end{thebibliography}
\end{document}